\documentclass[12pt]{article}
\usepackage{titlesec}
\usepackage{epsfig}
\usepackage{graphicx}
\usepackage{graphics}
\usepackage{amssymb}
\usepackage{amsmath}
\usepackage{amsfonts}
\usepackage{color}
\usepackage{textcomp}
\usepackage{latexsym}
\usepackage{cite}
\usepackage{titletoc}
\usepackage{tikz}
\usepackage{xcolor}
\usetikzlibrary{3d}
\usepackage{csquotes}
\usepackage{caption}
\usetikzlibrary{calc} 
\usetikzlibrary{angles,quotes}
\usepackage{csquotes}
\usepackage[title]{appendix}
\titleformat{\section}{\normalfont\Large\bfseries}{\thesection.}{1em}{}
\titlecontents{section}[0em]{\bfseries}{\thecontentslabel. }{}{\titlerule*[1pc]{.}\contentspage}
\usepackage[margin=2.5cm,footskip=0.6cm]{geometry}
\begin{document}

\begin{center}
{\Large{\bf Distribution Functions and Moments of Harmonically Trapped Inertial Run-and-Tumble Particles in a Shear-Thickening Medium}} \\
\ \\
by \\
Subhanker Howlader and Prasenjit Das\footnote{prasenjit.das@iisermohali.ac.in} \\
Department of Physical Sciences, Indian Institute of Science Education and Research -- Mohali, Knowledge City, Sector 81, SAS Nagar 140306, Punjab, India. \\
\end{center}
\begin{abstract}
\noindent 
We study the dynamics of an athermal inertial active particle confined in a harmonic potential and moving through shear-thickening media in one dimension. The activity is modeled using a symmetric dichotomous noise of strength $\Sigma$ and flipping rate $\lambda$, which mimics a run-and-tumble particle. The shear-thickening response of the medium is incorporated through a velocity-dependent viscous force of the form $g(v)\sim \tan(v)$. Starting from the Fokker-Planck equations corresponding to the particle’s dynamics, we numerically compute the time-dependent joint probability density function $f(x,v,t)$ using a \textit{finite-volume upwind scheme}. Next, we numerically solve the particle’s equations of motion to obtain the steady-state position $P_s(x)$ and velocity $W_s(v)$ distributions, and compare them with the marginal distributions obtained from $f(x,v,t)$ at sufficiently large $t$. The two approaches show excellent agreement. Both $P_s(x)$ and $W_s(v)$ exhibit several intriguing transitions as $\lambda$ is varied, and they eventually approach a Gaussian-like profile at sufficiently large $\lambda$. Our study of the time-dependent position-velocity correlation, $\langle xv \rangle(t)$, shows that larger values of $\lambda$ lead to faster relaxation to the steady-state when all other parameters are constant.
\end{abstract}

\newpage
\section{Introduction}\label{sec1} Over the past few decades, active materials have emerged as a new frontier in the field of soft matter~\cite{srarcm10,crhcggr16,bbjcp19}. They consist of an assembly of particles that harvest energy from their surroundings or convert their internal energy to generate directed motion. Therefore, they operate persistently out of equilibrium, violating detailed balance at the microscale. As a result, active matter systems show a wide variety of nonequilibrium behavior~\cite{mjstjmr13,jcmjc22}. Examples include collective motion, swarming, and motility-induced phase separation, which range from microbial communities to macroscopic patterns in ant colonies, fish schools, and bird flocks~\cite{nhpdkme14,mdbdmyss16,yfhyxzpmvk22,jkpsvvdg02,Fpgfgnp18,cwkyswy29}. Active particles also exhibit anomalous transport, including superdiffusion in run-and-tumble bacteria and active Brownian particles. Beyond these fundamental phenomena, the unique properties of active systems have many technological applications. Examples include bioinspired platforms and engineered systems, such as Janus particles, microrobots, and chemically or magnetically driven particles~\cite{pznnan13,ossjras17,zgglang17,llzacr19,asomasundar21,dsbsstnc22}. In medicine, notable applications are targeted drug delivery using light-driven and ultrasound-driven micromotors, as well as emerging strategies to combat bacterial infections and cancer~\cite{hyyhmyc20,jlccmmcdf22,msoldfuk22,qcxlyttlr24}.

Several theoretical models have been proposed to investigate the dynamics of active particles at the microscopic level. In most physical situations, active particles are small in mass and move in a dense fluid; therefore, their motion is considered to be in the overdamped regime, and thermal noise from the fluid medium plays a significant role. In this regime, three extensively studied models are the active Brownian particle (ABP)\cite{bmrspre18,fmjrtblcp23}, the active Ornstein–Uhlenbeck particle (AOUP)~\cite{mbcfn21,nwljpcghp22}, and the run-and-tumble particle (RTP)\cite{mjkksmr18,crzzprett23}. ABPs exhibit persistent random motion due to a self-propulsion force of constant magnitude, while their orientation undergoes rotational diffusion, typically modeled as Gaussian white noise. AOUPs are a variant of ABPs, where the orientational dynamics are replaced by a colored noise, known as the Ornstein–Uhlenbeck process. In contrast, RTPs alternate between \textit{run-state}, during which they move in a straight line at constant speed $v_{0}$, and a \textit{tumble-state}, where they undergo sudden reorientations of active force at a rate $\lambda$.

Recently, it has been observed that retaining the inertial term in the particle’s equation of motion -- referred to as inertial active matter (IAM) or active Langevin dynamics -- can substantially modify both single-particle statistics~\cite{cmjumbcp21,sclwlchl23,nwljpcghp22,mspresaho22} and collective behavior~\cite{dbg20,lowen20,rev1,rev2}. Natural examples of IAM range from large flying objects such as flamingos to macroscopic swimmers such as whales, while artificial examples are macrorobots, driven granular matter, etc. In most of the previous studies on IAM, the motion of particles in a Newtonian medium has been considered. In such media, the viscous force varies linearly with the particle’s speed. Owing to this linearity, closed-form expressions for the steady-state distributions of position and velocity for inertial RTPs (IRTPs) can be derived analytically. In reality, however, active particles often move through more complex environments -- such as polymer solutions, mucus, cellular fluids, and blood -- which are non-Newtonian, where the viscous force depends nonlinearly on the particle’s speed~\cite{esjbcssk14,nsrflgg19,lavmcrc23,jcpacwwkcfp23}. Motivated by the relevance of such environments, several studies have investigated the dynamics of active particles subjected to nonlinear dissipation~\cite{new0,new0a}. For example, Arsha and Sahoo theoretically studied the dynamics of an inertial active Ornstein--Uhlenbeck particle in the presence of nonlinear velocity-dependent friction~\cite{new1}. Similarly, Antonov \textit{et al.} investigated the dynamics of inertial active matter under Coulomb friction through both theoretical analysis and experiments~\cite{new2,new3}. Recently, Mondal and Das studied the dynamics of a free IRTP moving in a shear-thinning medium, modeling the viscous force as $g(v) \sim \tanh(v)$, and analytically obtained the steady-state velocity distribution $W_s(v)$~\cite{smpdaspf25}. Subsequently, Howlader \textit{et al.} extended this study to a shear-thickening medium by modeling the viscous force as $g(v) \sim \tan(v)$~\cite{showsmpdas25}. However, due to the nonlinear form of $g(v)$, they were unable to derive closed-form expressions for the position and the joint position-velocity probability density functions, even in the steady state.

In this paper, we focus on understanding the nonequilibrium steady state of an IRTP moving in a shear thickening medium under harmonic confinement in one dimension ($d=1$). The motivation for this study stems from the fact that active matter systems exhibit a variety of intriguing phenomena under confinement, leading to rich nonequilibrium behavior that is absent in unconfined systems. For example, the persistent interaction of active particles with physical boundaries leads to wall accumulation and boundary-layer formation~\cite{cfleenjp13,adjvspre18,rev3,rev4}. In this context, the morphology of bacterial colonies can be controlled by the geometry of the confining boundaries~\cite{elhwregpnas14,hwelregnjp16}. Similarly, clustering and motility-induced phase separation can be manipulated using different confinement geometries too, giving rise to a rich variety of nonequilibrium steady states~\cite{fmjrtblcpr23}. Harmonic confinement, which is the focus of our study, has an experimental realization in the controlled manipulation of microorganisms using optical tweezers. Such techniques have important applications in biosensing and controlled drug delivery~\cite{ayadpkyd,tweezer_book,rev5}. Recently, Buttinoni \textit{et al.} experimentally observed a transition from a unimodal to a bimodal position distribution of active Janus colloids confined in harmonic optical potentials under overdamped condition. They showed that this transition can be induced by increasing the self-propulsion velocity at fixed trap strength~\cite{buttinoni2022}.

Although the dynamics of an IRTP in a Newtonian medium under harmonic confinement have been studied previously, its behavior in a shear thickening medium remains largely unknown~\cite{dasbasupre24,majumder2026}. Dutta \textit{et al.} identify two dynamical regimes -- overdamped and underdamped -- arising from the competition of viscous, inertial, and activity timescales. Their study analytically characterizes the bounds and shape transitions of the steady state position and velocity distributions across different activity regimes. Most notably, they show that inertia produces multipeaked distribution functions in the highly active regime and smooth Gaussian-like distributions in the passive limit~\cite{dasbasupre24}. In contrast, our study considers dynamics in a shear-thickening medium where viscosity depends on velocity; we therefore investigate how the singularities of steady-state position and velocity distributions observed in Newtonian medium changes in shear-thickening medium, as well as their moments, depend on the flipping rate $\lambda$ of the active force. Furthermore, at sufficiently large $\lambda$, the particle is expected to fluctuate only in the vicinity of the center of the harmonic trap, thereby reducing the influence of confinement. In this regime, we anticipate that the steady-state velocity distribution $W_s(v)$ will exhibit transitions analogous to those observed for a free run-and-tumble particle.

We organize the paper in the following order. We present the details of the model and theoretical framework in Sec.~\ref{sec2}. Section~\ref{sec3} describes the numerical simulations and discusses the results. Finally, we summarize our results in Sec.~\ref{sec4}.

\section{Modeling and Theory}\label{sec2}
Consider the one-dimensional motion of an athermal IRTP moving through a shear-thickening medium confined in a harmonic trap. We model the viscous force experienced by the particle as $g(v)=A\tan(v/v_0)$, where $A$ is the amplitude of the viscous force and the $\tan(v/v_0)$ is a nonlinear function describing shear thickening nature of the medium with $v$ be the particle velocity and $v_0$ is a constant defining the characteristic velocity scale~\cite{menzel2015}. The equation of motion of the particle is given by
\begin{eqnarray}
\label{apeqn1}
\frac{dx}{dt} &=& v,\ \\
\label{apeqn2}
m \frac{dv}{dt} &=& -A\tan\left(\frac{v}{v_0}\right) - kx+ \xi(t).
\end{eqnarray}
In Eq.~(\ref{apeqn1}), $x$ denotes the particle’s position, while in Eq.~(\ref{apeqn2}), $m$ and $k$ represent the particle’s mass and the stiffness constant of the harmonic confinement, respectively. The term 
$\xi(t)$ represents the active force, modeled as a symmetric dichotomous noise\cite{smpdaspf25,showsmpdas25}. This noise takes two discrete values, $\pm\Delta$, switching between them with probability $\Gamma dt$ in time interval $dt$, where $\Gamma$ is the uniform transition rate. The mean value and autocorrelation of $\xi(t)$ are given by
\begin{eqnarray}  
\label{apeqn3}  
\langle \xi(t)  \rangle =0\quad\text{and}\quad\langle \xi(t) \xi(t^{\prime}) \rangle = \Delta^2 \exp\left(-2 \Gamma |t - t^{\prime}|\right).  
\end{eqnarray}
This noise effectively models the run-and-tumble dynamics of active particles in $d=1$.

Next, we nondimensionalize Eqs.~(\ref{apeqn1}) and (\ref{apeqn2}) by introducing the following rescaled variables for position, velocity, time and stiffness constant:
\begin{eqnarray}  
\label{apeqn4}  
x = \frac{mv_0^2}{A}\tilde{x}, \quad v = v_0 \tilde{v}, \quad  t = \frac{m v_0}{A} \tilde{t}, \quad \text{and} \quad k=\frac{A^2}{mv_0^2}\tilde{k},  
\end{eqnarray}
where the variables with tilde represent the corresponding dimensionless quantities. Substituting these into Eqs.~(\ref{apeqn1}) and (\ref{apeqn2}) and dropping the tildes, we get the nondimensional forms as  
\begin{eqnarray}
\label{apeqn5}
\frac{dx}{dt} &=& v,\ \\
\label{apeqn6}  
\frac{dv}{dt} &=& -\tan(v) - kx + \eta(t).  
\end{eqnarray}
Here, $\eta(t)$ is the scaled active force. It satisfies the following properties:  
\begin{eqnarray}  
\label{apeqn7}  
\langle \eta(t) \rangle = 0  \quad \text{and} \quad \langle \eta(t) \eta(t^{\prime}) \rangle = \Sigma^2 \exp\left(-2\lambda|t - t^{\prime}|\right),  
\end{eqnarray}  
where $\Sigma = \Delta/A$ and $\lambda = mv_0\Gamma/A$, respectively, represent the scaled strength and uniform flipping rate of the active force. 

%Table~\ref{table} lists the chosen forms of $g(v)$ used to describe different media.
%\begin{table}[ht]
%    \centering
%    \renewcommand{\arraystretch}{1.6} % Increase the row spacing
%    \caption{Choice of $g(v)$ corresponding to different media.}
%    \label{table}
%    \begin{tabular}{| c | c | c |}
%        \hline  
%        Newtonian & Shear thinning & Shear thickening  \\ [1ex] \hline
%        $g(v)=v$ & $g(v)=\tanh(v)$ & $g(v)=\tan(v)$ \\ \hline
%    \end{tabular}
%\end{table}

We write the Fokker-Planck~(FP) equations corresponding to the particle's equations of motion~(\ref{apeqn5}) and (\ref{apeqn6})~\cite{vbalakrishnan08}. We define $f_{\pm}(x, v, t)$ as the conditional joint probability density for the particle to have position $x$, velocity $v$ at time $t$, given the active force $\pm\Sigma$. The FP equations for these probability densities are as follows
\begin{eqnarray}
\label{apeqn8}
\frac{\partial}{\partial t} f_{+}(x, v, t) = -\frac{\partial}{\partial x} vf_{+}(x, v, t) + \frac{\partial}{\partial v} \left[\tan(v) + kx - \Sigma\right]f_{+}(x, v, t) + \lambda\left[f_{-}(x, v, t) - f_{+}(x, v, t)\right],\\
\label{apeqn9}
\frac{\partial}{\partial t} f_{-}(x, v, t) = -\frac{\partial}{\partial x} vf_{-}(x, v, t) + \frac{\partial}{\partial v} \left[\tan(v) + kx + \Sigma\right]f_{-}(x, v, t) + \lambda\left[f_{+}(x, v, t) - f_{-}(x, v, t)\right].
\end{eqnarray}
In Eqs.~(\ref{apeqn8}) and (\ref{apeqn9}), the first two terms on the right-hand side represent the deterministic evolution of the conditional probability densities in phase space~($x$-$v$ space) due to drift, the friction force $-\tan(v)$, restoring force $-kx$ and active force $+\Sigma$ or $-\Sigma$. The last term accounts for the flipping of the active force, which redistributes the probability densities between the two force states at a fixed position $x$ and velocity $v$. 

We define $f(x,v,t)$ be the total probability density of finding the particle at position $x$ with velocity $v$ at time $t$, irrespective of the state of the active force. It is given by
\begin{eqnarray}
\label{apeqn10}
f(x,v,t)=f_{+}(x,v,t)+f_{-}(x,v,t).
\end{eqnarray}
Since $\tan(v)$ is a nonlinear function, obtaining a full time-dependent solution of $f(x,v,t)$ or its steady state solution $f_s(x,v)$ in closed form is not feasible. Therefore, we compute $f(x,v,t)$ or $f_s(x,v)$ by numerically solving the FP equations (\ref{apeqn8}) and (\ref{apeqn9}). This enables us to validate results for various statistical quantities, such as the moments of $x$ and $v$, against the corresponding results obtained from independent particle-based simulations. In this paper, we refer to the calculation of various statistical quantities from $f(x,v,t)$, obtained by numerically solving the FP equations, as the \textit{semi-analytical method}.

Using $f(x,v,t)$, the time-dependent marginal position distribution function, $P(x,t)$, is obtained by integrating the joint distribution $f(x,v,t)$ over all velocities:
\begin{eqnarray}
\label{apeqn10a}
P(x,t) = \int dv\, f(x,v,t).
\end{eqnarray}
Similarly, the marginal velocity distribution function, $W(v,t)$, is obtained by integrating $f(x,v,t)$ over all positions:
\begin{eqnarray}
\label{apeqn10b}
W(v,t) = \int dx\, f(x,v,t).
\end{eqnarray}
In the steady state, we denote these distributions by $P_s(x)$ and $W_s(v)$, respectively. The average of any quantity $S(x,v)$ that depends on position and velocity can then be evaluated as
\begin{eqnarray}
\label{apeqn11}
\langle S(x,v)\rangle(t)=\int dx\int dv\,S(x,v)f(x,v,t).
\end{eqnarray}
Alternatively, $P(x,t)$, $W(v,t)$, and $\langle S(x,v)\rangle(t)$ can be computed directly from the particle trajectories $x(t)$ and $v(t)$ by solving the dynamical equations (\ref{apeqn5}) and (\ref{apeqn6}); we refer to this approach as the \textit{numerical method}. In this study, we compare the results obtained from both methods for various quantities.

\section{Numerical Simulation and Results}\label{sec3}
We numerically solve Eqs.~(\ref{apeqn5}) and (\ref{apeqn6}) using the Euler-discretization method to update the particle's position and velocity. The discretized version of the equations are as follows:
\begin{eqnarray}
\label{apeqn12}
x(t+dt)&=&x(t) + v(t) dt,\\
\label{apeqn13}
v(t+dt)&=&v(t) + [-\tan(v) - kx + \eta(t)] dt.
\end{eqnarray}
Here, $dt$ is the discretized time step. To generate the dichotomous noise $\eta(t)$, we follow the algorithm outlined in Ref.~\cite{ckeklpre06}. The probability $p$ that $\eta(t)$, starting from state $+\Sigma$(or $-\Sigma$), flips to state $-\Sigma$(or $+\Sigma$) in time interval $dt$ is given by
\begin{eqnarray}
\label{apeqn14}
p=\frac{1}{2}\{1-\exp(-2\lambda dt)\}.
\end{eqnarray}
Next, we generate a random number $w\in [0,1]$ using a uniform random number generator. When $p>w$, $\eta(t)$ flips. In our numerical simulations, we use $dt=0.001$. At $t=0$, we begin by placing all the particles at the origin with zero velocity. All the distribution functions and statistical results are obtained by using $10^6$ particles.

\begin{figure}[ht!]
\centering
\includegraphics*[width=0.50\textwidth]{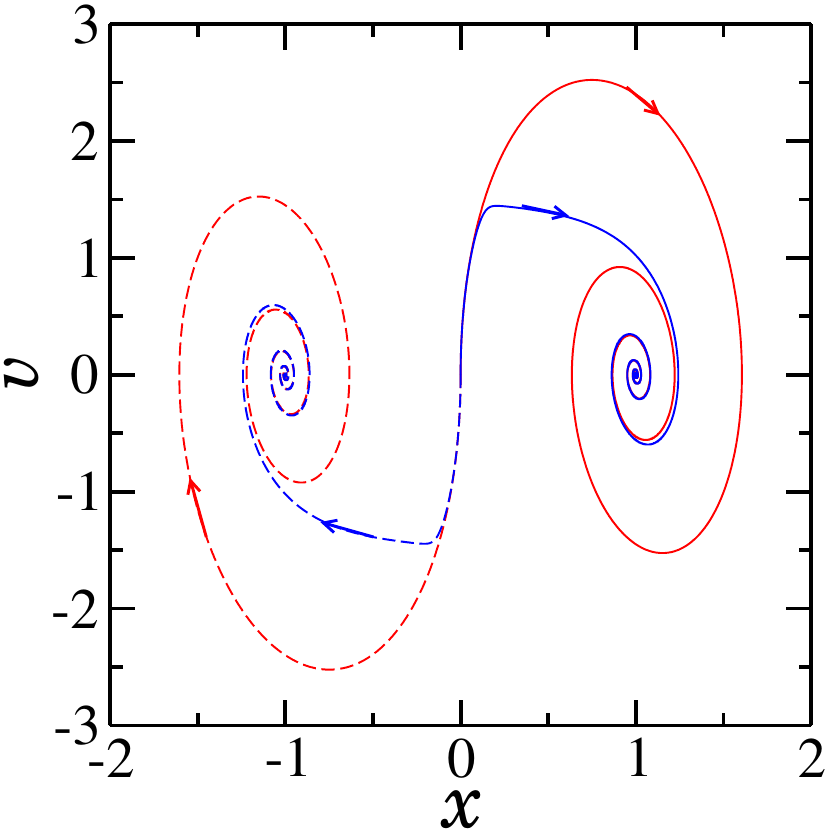}
\caption{\label{fig1} Phase-space trajectories of an IRTP in a harmonic potential with parameters $k = 10$ and $\lambda = 0$, i.e., under a fixed active force. The particle starts from the origin $(x(0), v(0)) = (0,0)$ of the $x$-$v$ plane. Solid lines correspond to the active force $\Sigma=+10$, while dashed lines correspond to $\Sigma=-10$. Red curves show trajectories for the Newtonian medium and blue curves for the shear thickening medium. Arrows indicate the direction of increasing time.}
\end{figure}
Figure~\ref{fig1} shows the phase-space trajectories of the IRTP for constant active forces $\eta(t)=+\Sigma$ and $\eta(t)=-\Sigma$ in the presence of viscous forces $-v$ and $-\tan(v)$, representing Newtonian and shear-thickening media, respectively. A constant $\eta(t)$ corresponds to $\lambda=0$. We set $k=10$ and $\Sigma=10$. Clearly, the dynamics exhibits two attractors located at ($\pm x_a$, 0), independent of the nature of the medium. However, the approach to these attractors depends on the nature of the medium. The relation between $x_a$ and the model parameters can be obtained from Eq.~(\ref{apeqn6}) and is given by $x_a=\Sigma/k$. Additionally, the phase-space area covered by a trajectory is larger for the Newtonian medium compared to the shear-thickening medium for the same amplitude of dissipation constant (unity in the scaled unit). This is because, the viscous force mimicking the shear-thickening medium bounds $v\in(-\pi/2, +\pi/2)$, whereas in the Newtonian case, no such bound exists, and the upper limits of $v$ are determined by both $\Sigma$ and $k$.

Next, we numerically solve the coupled FP equations~(\ref{apeqn8}) and (\ref{apeqn9}) using a \textit{finite-volume upwind scheme} to integrate them forward in time~\cite{pmpmcmhm05}. The details of the numerical scheme is provided in Appendix~\ref{appA}. At $t=0$, particles are assumed to be located at the origin $(0,0)$ of a two-dimensional $x$-$v$ space. The probability density functions $f_{+}$ and $f_{-}$ are then initialized as delta distributions centered at the origin. In all our simulations, we have assigned the values $f_{+}(0,0,0) = 0.2/\Delta x\Delta v$ and $f_{-}(0,0,0) = 0.8/\Delta x\Delta v$, respectively, so that the total probability is normalized. Here, $\Delta x$ and $\Delta v$ denote the discretized mesh sizes in $x$-$v$ space. Any positive values of $f_{+}(0,0,0)$ and $f_{-}(0,0,0)$ may be chosen, provided that their sum yields a normalized total probability. We consider a regular two-dimensional grid of size $2600\times 2000$ for $k=10$ and $\Sigma=10$. The grid extends from $-1.3$ to $+1.3$ in the $x$-direction and from $-1.57$ to $+1.57$ in the $v$-direction. The choice of $x \in [-1.3,,1.3]$ stems from the particle-based simulation data shown in Fig.~\ref{fig1}. To ensure probability conservation, we employ the no-flux boundary conditions, which keep $f_{\pm}$ zero (or negligibly small) at the boundaries throughout our simulation time. The discretized time step size is $\Delta t=3\times10^{-6}$, which provides sufficient numerical stability.
\begin{figure}[ht!]
\centering
\includegraphics*[width=0.999\textwidth]{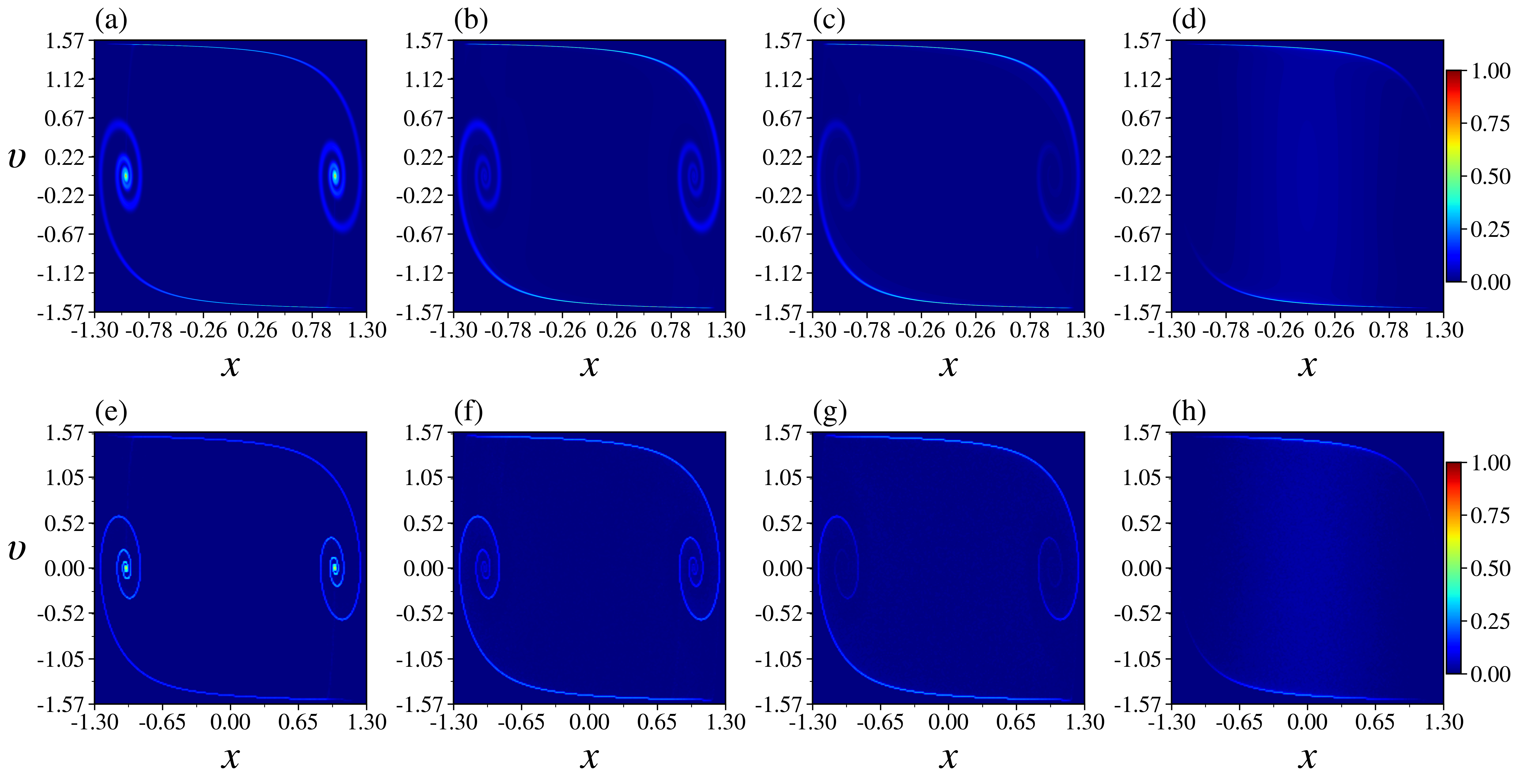}
\caption{\label{fig2} Density plots of the normalized steady-state probability distribution $f_s(x,v)$ in the $x$-$v$ space for $\Sigma = 10$, $k=10$, and different values of $\lambda$: (a, e) $\lambda = 0.2$, (b, f) $\lambda = 0.9$, (c, g) $\lambda = 1.7$, and (d, h) $\lambda = 7$. The top row [(a)–(d)] shows results obtained from the numerical solution of the FP equations, while the bottom row [(e)–(h)] shows results from particle-based simulations. The color bar in each subfigure indicates the corresponding values of $f_s(x,v)$ at each point.}
\end{figure}

At sufficiently large $t$, $f(x,v,t)$ approaches a steady state distribution $f_s(x,v)$. Figure~\ref{fig2} shows the density plots of the normalized $f_s(x,v)$, obtained via semi-analytical (top row) and numerical (bottom row) methods for different values of $\lambda$, as mentioned. In the numerical approach, $f_s(x,v)$ is computed using bin sizes $\Delta x=0.001$ and $\Delta v=0.005$ in the $x$-$v$ space. We observe a clear quantitative agreement between the results from both methods for the same set of parameters. For $\lambda=0.2$, the phase-space trajectories are mostly concentrated around the fixed points of the dynamics ($\pm x_a$, 0) in phase space, as shown in Figs.~\ref{fig2}(a) and \ref{fig2}(e). Upon increasing $\lambda$ to 0.9, the probability of finding the particle near ($\pm x_a$, 0) decreases, as evidenced by the reduced intensity of the density plots; compare Figs.~\ref{fig2}(a) and \ref{fig2}(b) (or Figs.~\ref{fig2}(e) and \ref{fig2}(f)), corresponding to $\lambda=0.2$ and $\lambda=0.9$, respectively. Since the total probability is conserved, the reduction in probability density near ($\pm x_a$, 0) is accompanied by a broadening of $f_s(x,v)$ around these points. As $\lambda$ increases further to 1.7, the higher flipping of $\eta(t)$ prevents the particles from reaching ($\pm x_a$, 0). Consequently, $f_s(x,v)$ spreads out in the vicinity of these points, as shown in Fig.~\ref{fig2}(c) and \ref{fig2}(g). For relatively large flipping rates, such as $\lambda=7$, the particles primarily fluctuate around $x=0$, with a higher probability of having velocities close to their maximum values [Figs.~\ref{fig2}(d, h)]. Thus, $f_s(x,v)$ exhibits several intriguing transitions as $\lambda$ is varied for fixed $\Sigma$ and $k$. The small quantitative differences between the results obtained from the FP equation [Figs.~\ref{fig2}(a)–(d)] and the particle trajectories [Figs.~\ref{fig2}(e)–(h)] arise primarily from finite sampling in the particle simulations and the different numerical discretization schemes employed in the two approaches. The distributions obtained from the particle trajectories are reconstructed from a finite ensemble of trajectories using histogram binning, whereas the FP equation directly yields the ensemble-averaged distribution function. Nevertheless, the excellent agreement in the spiral structures, peak locations, and extent of the distributions confirms the consistency of the two methods.
\begin{figure}
\centering
\includegraphics*[width=0.999\textwidth]{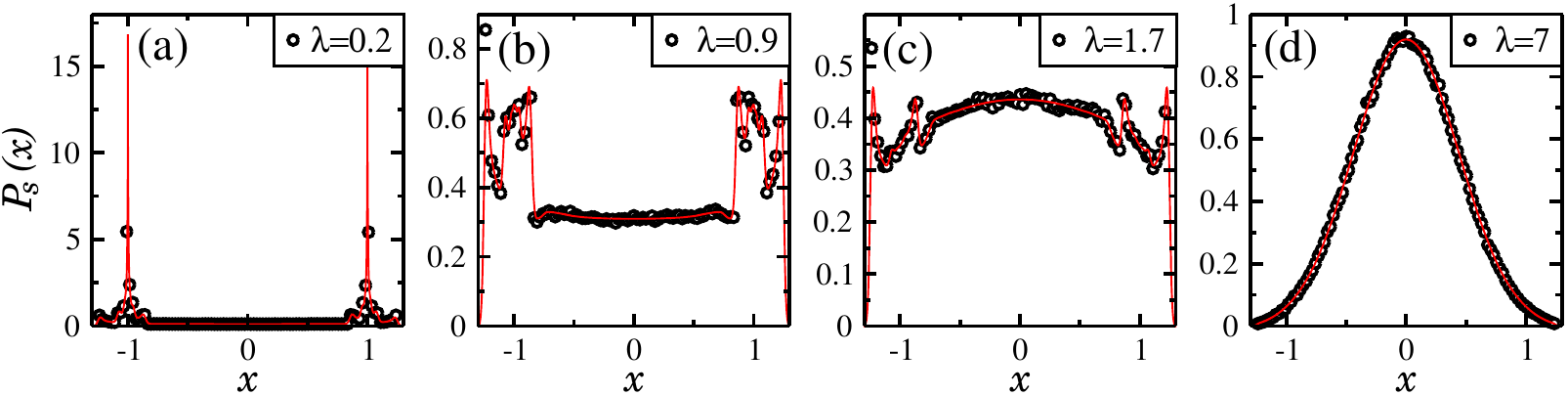}
\caption{\label{fig3} Plot of $P_s(x)$ for $\Sigma = 10$, $k = 10$, and various values of $\lambda$: (a) $\lambda=0.2$, (b) $\lambda=0.9$, (c) $\lambda=1.7$, and (d) $\lambda=7$, as indicated. Data obtained from the numerical method are shown by points, while the solid red lines represent the semi-analytical results.}
\end{figure}

To gain deeper insight into the transitions observed in $f_s(x,v)$ in Fig.~\ref{fig2}, we focus on the normalized marginal position and velocity distribution functions, $P_s(x)$ and $W_s(v)$. Figure~\ref{fig3} shows the plot of $P_s(x)$ for different values of $\lambda$. Clearly, the semi-analytical results obtained from Eq.~(\ref{apeqn10a}) show excellent agreement with the numerical data in all cases. For a small flipping rate, e.g., $\lambda=0.2$, $P_s(x)$ exhibits two distinct peaks at $x=\pm 1$, as shown in Fig.~\ref{fig3}(a). In this regime, most of the particles can reach the fixed points ($\pm x_a$,0) in the phase space before $\eta(t)$ reverses. Since we set $\Sigma=10$ and $k=10$, the expected locations of the peaks are $\pm\Sigma/k=\pm 1$, consistent with our observations. As $\lambda$ increases to 0.9, particles no longer have enough time to reach $x_a = \pm 1$ between successive reversals of $\eta(t)$. Consequently, the peaks around $x_a = \pm 1$ broaden due to the contribution from the spiral-like trajectories [see Fig.~\ref{fig1}] in phase space around $x_a=\pm1$ and $P_s(x)$ develops a nonzero probability density around $x = 0$, as shown in Fig.~\ref{fig3}(b). For an even higher flipping rate, $\lambda = 1.7$, the frequent reversals of $\eta(t)$ further enhance the probability near $x = 0$ while reducing the weight near $x = \pm 1$, as illustrated in Fig.~\ref{fig3}(c). For sufficiently large flipping rates, such as $\lambda=7$, particles predominantly fluctuate around $x=0$, resulting in a bell-shaped distribution centered at $x=0$, as shown in Fig.~\ref{fig3}(d). Increasing $\lambda$ further maintains the qualitative bell-shape of $P_s(x)$ while narrowing its width. 

\begin{figure}
\centering
\includegraphics*[width=0.99\textwidth]{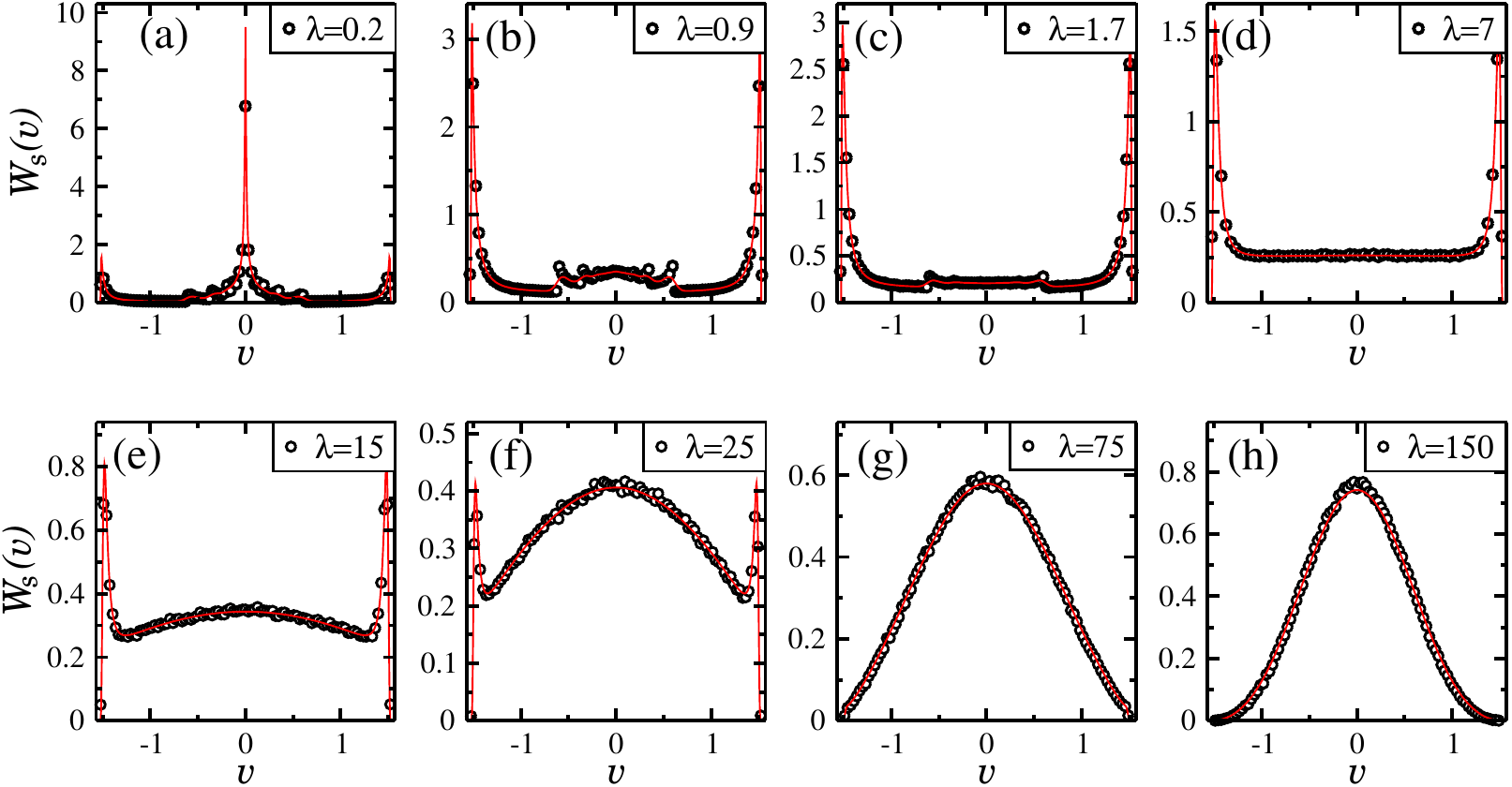}
\caption{\label{fig4} Plot of $W_s(v)$ for different values of $\lambda$: (a) $\lambda=0.2$, (b) $\lambda=0.9$, (c) $\lambda=1.7$, (d) $\lambda=7$, (e) $\lambda=15$, (f) $\lambda=25$, (g) $\lambda=75$, and (h) $\lambda=150$, as indicated. All other details are the same as those mentioned in the caption of Fig.~\ref{fig3}.}
\end{figure}

Figure~\ref{fig4} shows the plots of $W_s(v)$ for different values of $\lambda$. Our numerical data exhibit excellent agreement with the semi-analytical results obtained from Eq.~(\ref{apeqn10b}). For $\lambda=0$, all particles relax to the stable fixed points $(\pm 1, 0)$ in phase space, and therefore $W_s(v)$ becomes a delta function at $v=0$. For a small flipping rate, such as $\lambda=0.2$, a large fraction of particles remain near these fixed points. As a result, $W_s(v)$ maintains a pronounced central peak at $v=0$, as shown in Fig.~\ref{fig4}(a). At the same time, a smaller fraction of particles occasionally reach the maximum allowed velocities, giving rise to weak secondary peaks near the edges of the $v$ domain. As $\lambda$ increases to 0.9, particles continue to spend appreciable time near $x=\pm 1$ [see Fig.~\ref{fig3}(b)], but the frequent reversals of $\eta(t)$ impart finite velocities to them. Consequently, the central peak at $v=0$ diminishes, and the secondary peaks near the edges of the $v$ domain become more pronounced, as shown in Fig.~\ref{fig4}(b). For larger flipping rates, for example $\lambda=1.7$ [Fig.~\ref{fig4}(c)] and $\lambda=7$ [Fig.~\ref{fig4}(d)], the peaks near the edges of the $v$ domain decrease while the probability density around $v=0$ increases again. In this regime, the number of particles that can reach the maximum allowable velocities decreases with increasing $\lambda$, due to the more frequent flipping of $\eta(t)$.

Since, for $\lambda \geq 7$, particles fluctuate predominantly around $x = 0$, the influence of the harmonic potential on their motion becomes minimal. Our earlier study of the steady-state velocity distribution $Q_s(v)$ of free run-and-tumble particles moving in a shear-thickening medium revealed several intriguing transitions as the flipping rate $\lambda$ increases for a fixed $\Sigma$~\cite{showsmpdas25}: (a) For $\lambda < 1$, $Q_s(v)$ has a minimum at $v = 0$ and diverges at $v = \pm \tan^{-1}\Sigma$; (b) For $1 < \lambda < 1 + \Sigma^2$, a local maximum emerges at $v = 0$, minima appear at $v = \pm \tan^{-1}\sqrt{\lambda - 1}$, and $Q_s(v)$ continues to diverge at $v = \pm \tan^{-1}\Sigma$; (c) For $1 + \Sigma^2 < \lambda < 2(1 + \Sigma^2)$, $v = 0$ becomes the sole maximum, and $Q_s(v)$ approaches zero at $v = \pm \tan^{-1}\Sigma$ with a cusp-like profile; (d) For $\lambda > 2(1 + \Sigma^2)$, $Q_s(v)$ approaches zero at $v = \pm \tan^{-1}\Sigma$ without a cusp-like profile, while $v = 0$ remains the maximum. A natural question that arises is whether the transitions observed in the free-particle case will also appear -- at least qualitatively -- in $W_s(v)$ when $\lambda\geq 7$.

Figure~\ref{fig4}(d)-(h) shows  $W_s(v)$ at higher flipping rates $\lambda\geq7$. For $\lambda=7$, $W_s(v)$ exhibits a minimum at $v=0$ and diverges near $v \approx \pm \pi/2$ [Fig.~\ref{fig4}(d)], analogous to case (a) of $Q_s(v)$. For $\lambda=15$ and $\lambda=25$, the divergence at $v \approx \pm \pi/2$ persists, while a local maximum develops at $v=0$ [Figs.~\ref{fig4}(e)-(f)], consistent with case (b). The height of this central peak increases with $\lambda$. At $\lambda=75$, $W_s(v)$ displays a single peak at $v=0$, and the tails go to zero vertically, producing a cusp-like profile [Fig.~\ref{fig4}(g)], similar to case (c). Finally, for $\lambda = 150$, the single central peak remains, but the tails approach zero horizontally [Fig.~\ref{fig4}(h)], resembling case (d) of $Q_s(v)$. These observations indicates that the qualitative transitions previously identified in $Q_s(v)$ also appear in $W_s(v)$ at sufficiently large $\lambda$.

\begin{figure}[ht!]
\centering
\includegraphics*[width=0.90\textwidth]{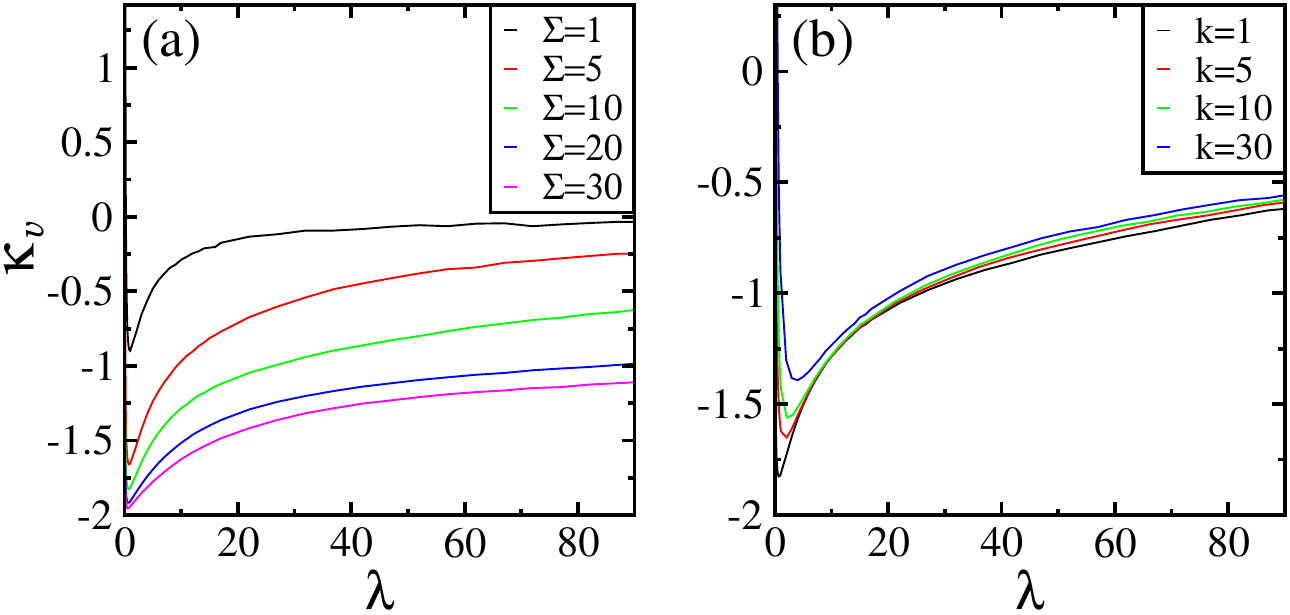}
\caption{\label{fig5} Plot of the kurtosis $\kappa_v$ of $W_s(v)$ as a function of $\lambda$ for (a) $k=1$ and different values of $\Sigma$ and (b) $\Sigma=10$ and various values of $k$, as indicated. These results are solely obtained from particle-based simulations.}
\end{figure}

Next, we characterize the shape of $W_s(v)$. As shown in Fig.~\ref{fig4}, $W_s(v)$ undergoes several intriguing transitions as $\lambda$ varies for fixed $\Sigma$, and eventually approaches a Gaussian-like form at large $\lambda$. To quantify the deviation from Gaussianity, we compute the kurtosis of $W_s(v)$, defined as
\begin{equation}
\label{apeqn15}
\kappa_v = \frac{\langle (v - \langle v \rangle)^4 \rangle}{\langle (v - \langle v \rangle)^2 \rangle^2} - 3.
\end{equation}
For the symmetric $\eta(t)$, $W_s(v)$ is symmetric about $v=0$. As a result, $\langle v\rangle=0$. So the expression in Eq.~(\ref{apeqn15}) simplifies to
\begin{equation}
\label{apeqn16}
\kappa_v = \frac{\langle v^4 \rangle}{\langle v^2 \rangle^2} - 3.
\end{equation}
A Gaussian distribution yields $\kappa_{v}=0$. Figure \ref{fig5} shows the variation of $\kappa_v$ with $\lambda$ for different values of $k$ and $\Sigma$, obtained from particle-based simulations. For a fixed $k = 1$, $\kappa_v$ approaches zero with increasing $\lambda$ more rapidly when $\Sigma$ is small, as shown in Fig.~\ref{fig5}(a). This is because, for small $\Sigma$ and sufficiently large $\lambda$, particles predominantly fluctuate around $x=0$ and $v=0$, resulting in a narrower, more Gaussian-like $W_s(v)$. In contrast, a larger $\Sigma$ drives particles to attain higher velocities even at the same $\lambda$, and consequently, $\kappa_v$ exhibits a larger deviation from zero. In Fig.~\ref{fig5}(b), we plot $\kappa_v$ as a function of $\lambda$ for different values of $k$ at fixed $\Sigma = 10$. For all $k$, $\kappa_v$ approaches zero with increasing $\lambda$, indicating that $W_s(v)$ approaches a Gaussian-like form in the large-$\lambda$ limit. The dependence on $k$ is relatively weak because, for sufficiently large $\lambda$, both $P_s(x)$ and $W_s(v)$ are dominated by fluctuations around $x=0$ and $v=0$, largely independent of $k$.

\begin{figure}
\centering
\includegraphics*[width=0.99\textwidth]{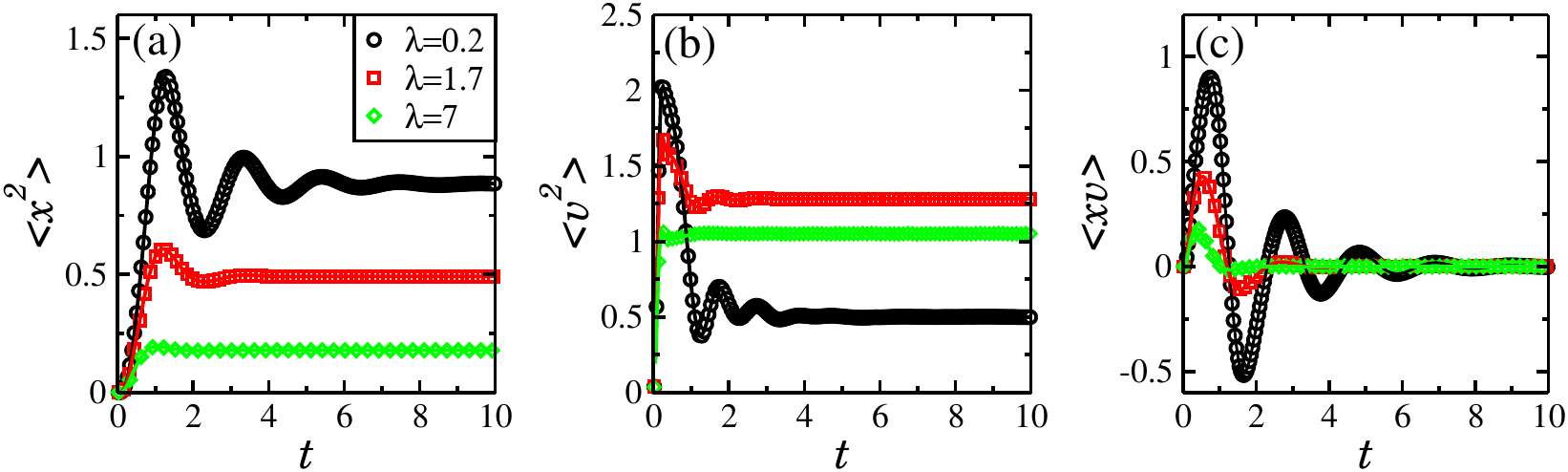}
\caption{\label{fig6}Plot of (a) $\langle x^2\rangle$ vs. $t$, (b) $\langle v^2\rangle$ vs. $t$, and (c) $\langle xv\rangle$ vs. $t$ for $\Sigma = 10$, $k=10$, and various values of $\lambda$, as indicated. The numerical results are shown by points, while the solid lines represent data obtained from the semi-analytical method.}
\end{figure}

Finally, we examine the time dependence of the moments of position and velocity. For a symmetric $\eta(t)$, both $\langle x \rangle$ and $\langle v \rangle$ become zero in the steady state (not shown here). Figure~\ref{fig6}(a) shows $\langle x^2\rangle$ as a function of $t$ for different values of $\lambda$ at fixed $\Sigma$ and $k$, as mentioned. In all cases, $\langle x^2\rangle$ saturates at long times, and its steady-state value decreases systematically with increasing $\lambda$. This is because, for small $\lambda$, particles spend more time away from $x=0$, whereas for large $\lambda$ they fluctuate predominantly around $x=0$. Figure~\ref{fig6}(b) shows the corresponding behavior of $\langle v^2\rangle$. Similar to $\langle x^2\rangle$, $\langle v^2\rangle$ attains a steady-state value, $\langle v^2\rangle_s$. However, its dependence on $\lambda$ is not monotonic. The smallest value of $\langle v^2\rangle_s$ occurs at $\lambda=0.2$, where $W_s(v)$ is sharply peaked at $v=0$ [Fig.~\ref{fig4}(a)]. For $\lambda=1.7$, $\langle v^2\rangle_s$ is larger than for $\lambda=7$, as $W_s(v)$ has more pronounced peaks near $v \approx \pm \pi/2$ at $\lambda=1.7$ [Fig.~\ref{fig4}(c)-(d)]. Next, Fig.~\ref{fig6}(c) shows the time evolution of $\langle xv\rangle$, which quantifies the correlation between position and velocity. In all cases, $\langle xv\rangle$ decays to zero at later times, indicating that $x$ and $v$ become uncorrelated in the steady state. The time required for this decay systematically decreases with increasing $\lambda$, as more frequent flippings of $\eta(t)$ accelerate relaxation to the steady state. Further, our numerical data show good agreement with the semi-analytical results in all cases.

\section{Summary}\label{sec4}
In this paper, we have studied the dynamics of an athermal inertial run-and-tumble particle moving in a shear-thickening medium in $d=1$. The particle is confined by a harmonic potential of the form $\tfrac{1}{2}kx^2$, where $k$ is the stiffness constant. A symmetric dichotomous noise, $\eta(t)$, that takes values $\pm\Sigma$ and switches between them at a uniform flipping rate $\lambda$ models the run-and-tumble dynamics of the particle. The shear-thickening nature of the medium is described by a velocity-dependent viscous force of the form $g(v)\sim \tan(v)$. As a result, the particle velocity always remains within the range $-\pi/2<v<\pi/2$.

We began by rescaling the equations of motion for the particle. In the absence of tumbling ($\lambda=0$), $\eta(t)$ takes constant values $\pm\Sigma$. Under this condition, the phase space ($x$-$v$ space) contains two fixed points located at ($x_a$,$v_a$)=($\pm\Sigma/k$,0). As a result, for small but nonzero $\lambda$, the steady-state position and velocity distributions, $P_s(x)$ and $W_s(v)$, show pronounced peaks near $x=x_a$ and $v=0$, respectively. As $\lambda$ increases, a large number of particles no longer have sufficient time to reach the fixed points due to frequent flipping of $\eta(t)$. Consequently, they attain nonzero velocities, causing $P_s(x)$ to develop broader peaks near $x=x_a$ and to acquire a finite value at $x=0$. At the same time, $W_s(v)$ develops peaks near $v\approx\pm\pi/2$. As we increase $\lambda$ further, particles predominantly fluctuate around $x=0$, and $P_s(x)$ gradually approaches a Gaussian-like profile. Once $P_s(x)$ becomes sharply peaked at $x=0$ for sufficiently large $\lambda$, the influence of the confining harmonic potential on the particle dynamics becomes minimal. In this limit, $W_s(v)$ exhibits the same sequence of transitions as those observed for a free run-and-tumble particle in a shear-thickening medium. In all cases, our numerical results agree well with the semi-analytical results. The complex interplay between the active force and the dissipative dynamics plays a crucial role in generating the observed non-Gaussian distributions.

We have characterized the shapes of $W_s(v)$ by computing its kurtosis $\kappa_v$ as a function of $\lambda$, taking $\Sigma$ and $k$ as control parameters. For a fixed $\Sigma$, $\kappa_v$ approaches zero as $\lambda$ increases for all values of $k$, showing only a weak dependence on $k$. In contrast, for a fixed $k$, $\kappa_v$ approaches zero more rapidly at smaller $\Sigma$ as $\lambda$ increases. Further, the time-dependent position-velocity correlation, $\langle xv \rangle(t)$, shows that larger values of $\lambda$ lead to faster relaxation to the steady-state when all other parameters are constant. We believe that the results presented above offer valuable insights into the position and velocity statistics of a harmonically trapped run-and-tumble particle in a shear-thickening medium.
\\ \ \\
\noindent{\bf Acknowledgments:} SH acknowledges financial support from IISER Mohali through a Senior Research Fellowship.
\\ \ \\
\noindent{\bf Conflict of interest:} The authors have no conflicts to disclose.

\appendix
\section{Numerical Solution of the Fokker–Planck Equations}\label{appA}
We use a finite-volume upwind scheme to numerically solve the FP equations. This scheme is known to preserve the conservation of total probability in phase space. The FP equations~\eqref{apeqn8} and \eqref{apeqn9} can be written as
\begin{equation}
\label{apeqn22}
\frac{\partial f_{\pm}}{\partial t} = -\frac{\partial}{\partial x}\Big(a(x,v),f_{\pm}\Big)
-\frac{\partial}{\partial v}\Big(b_{\pm}(x,v),f_{\pm}\Big) +S_{\pm},
\end{equation}
where $a(x,v)=v$, $b_{\pm}(x,v)=-[\tan(v)+kx\mp\Sigma]$, and $S_{\pm}=\lambda\left[f_{\mp}(x,v,t)-f_{\pm}(x,v,t)\right]$. Thus, our FP equations have the structure of an advection equation with a source term.

\begin{figure}
\centering
\includegraphics*[width=0.99\textwidth]{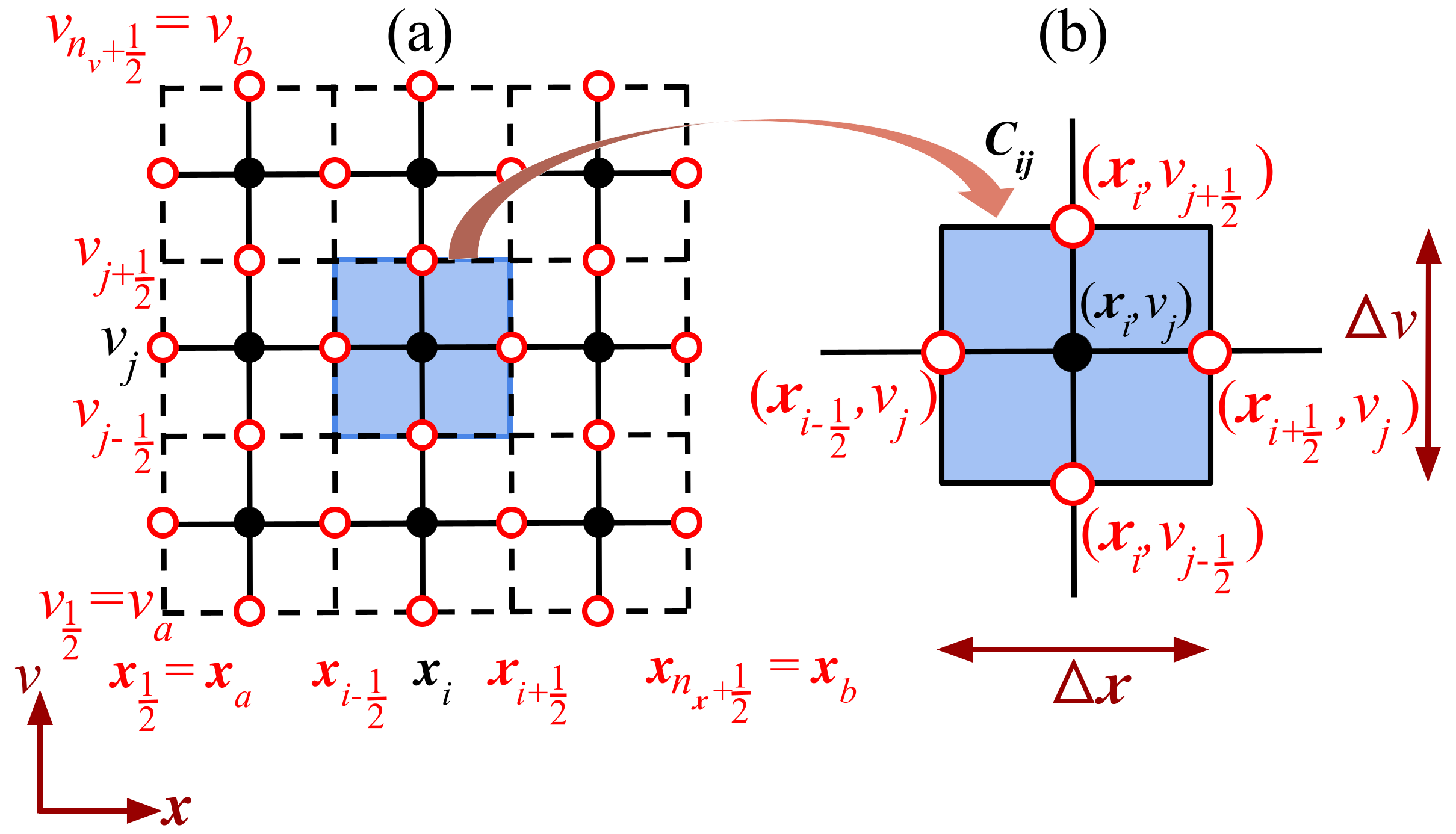}
\caption{\label{fig9} (a) Schematic representation of the discretization of the $x$-$v$ phase space into uniform control volumes. The black filled circles denote the cell centers $(x_i,v_j)$, while the red open circles denote the cell faces at $x_{i\pm1/2}$ and $v_{j\pm1/2}$. (b) Magnified view of the control volume $C_{ij}$ centered at $(x_i,v_j)$, showing its neighboring cell centers, face locations, and grid spacings $\Delta x$ and $\Delta v$.}
\end{figure}

We uniformly divide the position interval $x\in[x_a,x_b]$ and the velocity interval $v\in[v_a,v_b]$ into $n_x$ and $n_v$ grid points, respectively, with mesh sizes $\Delta x=(x_b-x_a)/n_x$ and $\Delta v=(v_b-v_a)/n_v$. The locations of the cell centers and faces are given by
\begin{eqnarray}
x\text{-space:}~~ x_i=x_a+\left(i-\frac12\right)\Delta x,\quad i=1,\ldots,n_x;
\quad
x_{i+\frac12}=x_a+i\,\Delta x,\quad i=0,\ldots,n_x,\\
v\text{-space:}~~v_j=v_a+\left(j-\frac12\right)\Delta v,\quad j=1,\ldots,n_v;
\quad
v_{j+\frac12}=v_a+j\,\Delta v,\quad j=0,\ldots,n_v.
\end{eqnarray} 
The area of each control volume in phase space ($x$-$v$ space) becomes $C_{ij}=\Delta x\times\Delta v$, where $\Delta x=x_{i+\frac12}-x_{i-\frac12}$ and $\Delta v=v_{j+\frac12}-v_{j-\frac12}$. Figure~\ref{fig9} shows the discretization of the phase space into uniform control volumes, with a magnified view of a representative cell $C_{ij}$. Next, by integrating Eq.~\eqref{apeqn22} over the cell $C_{ij}$ and the time interval $[t_n,t_{n+1}]$, with $\Delta t=t_{n+1}-t_n$, we obtain the following update equation:
\begin{equation}
\tilde{f}_{\pm,i,j}^{\,n+1}
=
\tilde{f}_{\pm,i,j}^{\,n}
-\frac{\Delta t}{\Delta x}
\left(
F_{\pm,i+\frac12,j}^{\,n}
-
F_{\pm,i-\frac12,j}^{\,n}
\right)
-\frac{\Delta t}{\Delta v}
\left(
G_{\pm,i,j+\frac12}^{\,n}
-
G_{\pm,i,j-\frac12}^{\,n}
\right)
+\Delta t\,S_{\pm,i,j}^{\,n},
\label{eq:fvm_update}
\end{equation}
where
\begin{equation}
\widetilde{f}_{\pm,i,j}^{\,n}
=
\frac{1}{\Delta x\,\Delta v}
\int_{x_{i-\frac12}}^{x_{i+\frac12}}
\int_{v_{j-\frac12}}^{v_{j+\frac12}}
f_{\pm}(x,v,t_n)\,dv\,dx,
\label{apeqn29}
\end{equation}
\begin{equation}
F_{\pm,i\pm\frac12,j}^{\,n}
=
\frac{1}{\Delta t\,\Delta v}
\int_{t_n}^{t_{n+1}}
\int_{v_{j-\frac12}}^{v_{j+\frac12}}
a(x_{i\pm \frac 12},v)f_{\pm}({x_{i\pm \frac 12}},v,t)\,dv\,dt ,
\label{apeqn34}
\end{equation}
\begin{equation}
G_{\pm,i,j\pm\frac12}^{\,n}
=
\frac{1}{\Delta t\,\Delta x}
\int_{t_n}^{t_{n+1}}
\int_{x_{i-\frac12}}^{x_{i+\frac12}}
b_\pm(x,v_{i\pm \frac 12})f_{\pm}(x,v_{i\pm \frac 12},t)\,dx\,dt,
\label{apeqn37}
\end{equation}
\begin{equation}
S_{\pm,i,j}^{\,n}
=
\frac{1}{\Delta t\,\Delta x\,\Delta v}
\int_{t_n}^{t_{n+1}}
\int_{x_{i-\frac12}}^{x_{i+\frac12}}
\int_{v_{j-\frac12}}^{v_{j+\frac12}}
S_{\pm}(x,v,t)\,dv\,dx\,dt.
\label{apeqn39}
\end{equation}
The details of the above scheme are presented in Ref.~\cite{pmpmcmhm05}.

We approximate $F_{\pm,i\pm\frac12,j}^{n}$  and $G_{\pm,i,j\pm\frac12}^{n}$ at the $x$- and $v$-cell interfaces, respectively, using an upwind scheme. Next, we apply an explicit method to the time integrals, leading to an explicit upwind scheme~\cite{hussain2020,umascher08,yyl2020,bradiaelhami}. To simplify the notation, we write $F_{\pm,i\pm\frac12,j}^{n}=F_{\pm,i\pm\frac12,j}$, $G_{\pm,i,j\pm\frac12}^{n}=G_{\pm,i,j\pm\frac12}$, $f_{\pm}(x_i,v_j,t_n)=f_{\pm,i,j}$, $S_{\pm}(x_i,v_j,t_n)=S_{\pm,i,j}$, $a(x_{i \pm\frac12},v_j)=a_{i\pm\frac12,j}$, and $b_{\pm}(x_i,v_{j\pm\frac12})=b_{\pm,i,j\pm\frac12}$. The selection of $F_{\pm,i\pm\frac12,j}^{n}$  and $G_{\pm,i,j\pm\frac12}^{n}$ under the explicit upwind scheme are
\begin{equation}
F_{\pm,i+\frac12,j}
=
\begin{cases}
a_{i+\frac12,j}f_{\pm,i,j}, & a_{i+\frac12,j}\ge 0,\\[4pt]
a_{i+\frac12,j}\,f_{\pm,i+1,j}, & a_{i+\frac12,j}<0,
\end{cases}
\end{equation}
\begin{equation}
F_{\pm,i-\frac12,j}
=
\begin{cases}
a_{i-\frac12,j}\,f_{\pm,i-1,j}, & a_{i-\frac12,j}\ge 0,\\[4pt]
a_{i-\frac12,j}\,f_{\pm,i,j}, & a_{i-\frac12,j}<0.
\end{cases}
\end{equation}
and
\begin{equation}
G_{\pm,i,j+\frac12}
=
\begin{cases}
b_{\pm,i,j+\frac12}\,f_{\pm,i,j}, & b_{\pm,i,j+\frac12}\ge 0,\\[4pt]
b_{\pm,i,j+\frac12}\,f_{\pm,i,j+1}, & b_{\pm,i,j+\frac12}<0,
\end{cases}
\end{equation}
\begin{equation}
G_{\pm,i,j-\frac12}
=
\begin{cases}
b_{\pm,i,j-\frac12}\,f_{\pm,i,j-1}, & b_{\pm,i,j-\frac12}\ge 0,\\[4pt]
b_{\pm,i,j-\frac12}\,f_{\pm,i,j}, & b_{\pm,i,j-\frac12}<0.
\end{cases}
\end{equation}
Further, we impose no-flux boundary conditions on all four sides of the computational domain, so that the fluxes through the outer cell faces vanish. A Taylor series  expansion around $(x_i,v_j)$, followed by an explicit method for the time integrals, gives $\widetilde{f}_{\pm,i,j}^{\,n}\approx f_\pm(x_i,v_j,t_n)=f_{\pm,i,j}$~\ and $S_{\pm,i,j}^{\,n}\approx S_\pm(x_i,v_j,t_n)=S_{\pm,i,j}$, respectively~\cite{yyuany19}.


\begin{thebibliography}{99}
\bibitem{srarcm10}S. Ramaswamy, Annu. Rev. Condens. Matter. Phys. \textbf{1}, 323 (2010).

\bibitem{crhcggr16}C. Bechinger, R. Di Leonardo, H. L\"{o}wen, C. Reichhardt, G. Volpe, and G. Volpe, Rev. Mod. Phys. \textbf{88}, 045006 (2016).

\bibitem{bbjcp19}E. W. Burkholder and J. F. Brady, J. Chem. Phys. \textbf{150}, 184901 (2019).

\bibitem{mjstjmr13}M. C. Marchetti, J. F. Joanny, S. Ramaswamy, T. B. Liverpool, J. Prost, M. Rao, and R. A. Simha, Rev. Mod. Phys. \textbf{85}, 1143 (2013).

\bibitem{jcmjc22}J. Tailleur, and others (eds.), \textit{Active Matter and Nonequilibrium Statistical Physics: Lecture Notes of the Les Houches Summer School: Volume 112, September 2018} (Oxford University Press, Oxford, 2022).

\bibitem{jkpsvvdg02}J. K. Parrish, S. V. Viscido, and D. Gr\"{u}nbaum, Biol. Bull. \textbf{202}, 296 (2002).

\bibitem{nhpdkme14}N. H. P. Nguyen, D. Klotsa, M. Engel, and S. C. Glotzer, Phys. Rev. Lett. \textbf{112}, 075701 (2014).

\bibitem{mdbdmyss16}M. Driscoll, B. Delmotte, M. Youssef,
S. Sacanna,  A. Donev, and  P. Chaikin, Nat. Phys. \textbf{13}, 375 (2016).

\bibitem{Fpgfgnp18}O. Feinerman, I. Pinkoviezky, A. Gelblum, E. Fonio, and N. S. Gov, Nat. Phys. \textbf{14}, 683 (2018).

\bibitem{cwkyswy29}D. Chen, Y. Wang, G. Wu, M. Kang, Y. Sun, and W. Yu, Chaos \textbf{29}, 113118 (2019).

\bibitem{yfhyxzpmvk22}Y. Fu, H. Yu, X. Zhang, P. Malgaretti, V. Kishore, and W. Wang, Micromachines \textbf{13}, 295 (2022).

\bibitem{pznnan13}K. E. Peyer, L. Zhang, and B. J. Nelson, Nanoscale \textbf{5}, 1259 (2013).

\bibitem{ossjras17}H. Oh, A. R. Shirazi, C. Sun, and Y. Jin, Robot. Auton. Syst. \textbf{91}, 83 (2017).

\bibitem{zgglang17}J. Zhang, B. A. Grzybowski, and S. Granick, Langmuir \textbf{33}, 6964 (2017).

\bibitem{llzacr19}T. Li, Y. Li, and T. Zhang, Acc. Chem. Res. \textbf{52}, 288 (2019).

\bibitem{asomasundar21}A. Somasundar and A. Sen, Small \textbf{17}, 2007102 (2021).

\bibitem{dsbsstnc22}S. R. Dabbagh, M. R. Sarabi, M. T. Birtek, S. Seyfi, M. Sitti, and S. Tasoglu, Nat. Commun. \textbf{13}, 5875 (2022).

\bibitem{hyyhmyc20}H. Y. Yoon, H. M. Yang, C. H. Kim, Y. T. Goo, M. J. Kang, S. Lee, and Y. W. Choi, Expert Opin. Drug Deliv. \textbf{17}, 1555 (2020).

\bibitem{jlccmmcdf22}J. Li, C. C. Mayorga-Martinez, C.-D. Ohl, and M. Pumera, Adv. Funct. Mater. \textbf{32}, 2102265 (2022).

\bibitem{msoldfuk22}M. S. Ozaydin, L. Doganturk, F. Ulucan-Karnak, O. Akdogan, and P. Erkoc, Prosthesis \textbf{4}, 424 (2022).

\bibitem{qcxlyttlr24}Q. Cheng, X. Lu, Y. Tai, T. Luo, and R. Yang,  ACS Biomater. Sci. Eng. \textbf{10}, 5562 (2024).

\bibitem{bmrspre18}U. Basu, S. N. Majumdar, A. Rosso, and G. Schehr, Phys. Rev. E \textbf{98}, 062121 (2018).

\bibitem{fmjrtblcp23}F. Moore, J. Russo, T. B. Liverpool, and C. P. Royall, J. Chem. Phys. \textbf{158}, 104907 (2023).

\bibitem{mbcfn21}D. Martin, J. O’Byrne, M. E. Cates, \'E. Fodor, C. Nardini, J. Tailleur, and F. van Wijland, Phys. Rev. E \textbf{103}, 032607 (2021).

\bibitem{nwljpcghp22}G. H. P. Nguyen, R. Wittmann, and H. L\"{o}wen, J. Phys.: Condens. Matter \textbf{34}, 035101 (2022).

\bibitem{mjkksmr18}K. Malakar, V. Jemseena, A. Kundu, K. V. Kumar, S.
Sabhapandit, S. N. Majumdar, S. Redner, and A. Dhar, J. Stat. Mech. \textbf{2018}, 043215 (2018).

\bibitem{crzzprett23}C. Roberts and Z. Zhen, Phys. Rev. E \textbf{108}, 014139 (2023).

\bibitem{cmjumbcp21}L. Caprini and U. M. B. Marconi, J. Chem. Phys. \textbf{154}, 024902 (2021).

\bibitem{mspresaho22}M. Muhsin and M. Sahoo, Phys. Rev. E \textbf{106}, 014605 (2022).

\bibitem{sclwlchl23}A. R. Sprenger, L. Caprini, H. L\"{o}wen, and R. Wittmann, J. Phys.: Condens. Matter \textbf{35}, 305101 (2023).

\bibitem{dbg20} C. Dai, I. R. Bruss, and S. C. Glotzer, Soft Matter \textbf{16}, 2847 (2020).

\bibitem{lowen20} H. L\"{o}wen, J. Chem. Phys. \textbf{152}, 040901 (2020).

\bibitem{rev1} L. Hecht, S. Mandal, H. L\"{o}wen, and B. Liebchen, Phys. Rev. Lett. \textbf{129}, 178001 (2022).

\bibitem{rev2} L. Caprini, D. Breoni, A. Ldov, C. Scholz, and H. L\"{o}wen, Commun. Phys. \textbf{7}, 343 (2024).

\bibitem{esjbcssk14}E. S. Vasquez, J. Bowser, C. Swiderski, K. B. Walters, and S. Kundu, RSC Advances \textbf{4}, 34780 (2014).

\bibitem{nsrflgg19}E. Nader, S. Skinner, M. Romana, R. Fort, N. Lemonne, N. Guillot, A. Gauthier, S. Antoine-Jonville, C. Renoux, M.-D. Hardy-Dessources, E. Stauffer, P. Joly, Y. Bertrand, and P. Connes,  Front Physiol. \textbf{10}, 1329 (2019).

\bibitem{lavmcrc23}K. S. M. Keithley, J. Palmerio, H. A. Escobedo IV, J. Bartlett, H. Huang, L. A. Villasmil, and M. Cromer, Rheol. Acta \textbf{62}, 377 (2023).

\bibitem{jcpacwwkcfp23}J. C. P. Hollister, A. C. Wang, W. Kim, C. C. Giza, M. L. Prins, and H. P. Kavehpour, Front. Phys. \textbf{11}, 1308136 (2023).

\bibitem{new0} P. M. Geffert and W. Just, Phys. Rev. E \textbf{95}, 062111 (2017).

\bibitem{new0a} Z. Zhang, S. Yuan, and S. Komura, New J. Phys. \textbf{26}, 093036 (2024).

\bibitem{new1} N. Arsha and M. Sahoo, Physica A \textbf{656}, 130202 (2024).

\bibitem{new2} A. P. Antonov, L. Caprini, A. Ldov, C. Scholz, and H. L\"{o}wen, Phys. Rev. Lett. \textbf{133}, 198301 (2024).

\bibitem{new3} A. P. Antonov, M. Musacchio, H. L\"{o}wen, and L. Caprini, Nat. Commun. \textbf{16}, 7235 (2025).

\bibitem{smpdaspf25}S. Mondal and P. Das,  Phys. Fluids \textbf{37}, 057114 (2025).

\bibitem{showsmpdas25}S. Howlader, S. Mondal, and P. Das, Phys. Rev. E \textbf{112}, 025403 (2025).

\bibitem{cfleenjp13}C. F. Lee, New J. Phys. \textbf{15}, 055007 (2013).

\bibitem{adjvspre18}A. Duzgun and J. V. Selinger, Phys. Rev. E \textbf{97}, 032606 (2018).

\bibitem{rev3} L. Caprini, and U. M. B. Marconi, Soft Matter \textbf{14}, 9044 (2018).

\bibitem{rev4} S. Das, G. Gompper, and R. G. Winkler, New J. Phys. \textbf{20}, 015001 (2018).

\bibitem{elhwregpnas14}E. Lushi, H. Wioland, and R. E. Goldstein, Proc. Natl. Acad. Sci. \textbf{111}, 9733 (2014).

\bibitem{hwelregnjp16}H. Wioland, E. Lushi, and R. E. Goldstein, New J. Phys. \textbf{18}, 075002 (2016).

\bibitem{fmjrtblcpr23}F. Moore, J. Russo, T. B. Liverpool, and C. P. Royall, J. Chem. Phys. \textbf{158}, 104907 (2023).

\bibitem{ayadpkyd}A. Yadav, A. Dutta, P. Kumar, Y. Dahan, A. Aranovich, and M. Feingold, Phys. Rev. E \textbf{101}, 062402 (2020).

\bibitem{tweezer_book}A. Ashkin, \textit{Optical Trapping and Manipulation of Neutral Particles Using Lasers: A Reprint Volume with Commentaries} (World Scientific, London, 2006).

\bibitem{rev5} E. A. Lisin, and I. I. Lisina, Phys. Rev. E \textbf{113}, 015417 (2026).

\bibitem{buttinoni2022} I. Buttinoni, L. Caprini, L. Alvarez, F. J. Schwarzendahl, and H. L\"{o}wen, Europhys. Lett. \textbf{140}, 27001 (2022).

\bibitem{dasbasupre24}D. Dutta, A. Kundu, S. Sabhapandit, and U. Basu, Phys. Rev. E \textbf{110}, 044107 (2024).

\bibitem{majumder2026} M. Gu\'{e}neau, S. N. Majumdar, and G. Schehr, arXiv:2602.08436 (2026).

\bibitem{menzel2015} A. M. Menzel, Phys. Rev. E \textbf{92}, 052302 (2015).

\bibitem{vbalakrishnan08}V. Balakrishnan, \textit{Elements of Nonequilibrium Statistical Mechanics} (Ane Books, New Delhi, 2008).

\bibitem{ckeklpre06}C. Kim and E. K. Lee, Phys. Rev. E \textbf{73}, 026101 (2006).

\bibitem{pmpmcmhm05}P. Majumdar, \textit{Computational Methods for Heat and Mass Transfer, first ed.} (Taylor and Francis, New York, 2005).

\bibitem{hussain2020} A. Hussain, Z. Zheng, and E. F. Anley, Mathematics \textbf{8}, 1869 (2020).

\bibitem{umascher08} U. M. Ascher, \textit{Numerical Methods for Evolutionary Differential Equations} (SIAM, Philadelphia, PA, USA, 2008).

\bibitem{yyl2020} Y. Du, Y. Wang, and L. Yuan, Appl. Math. \textbf{39}, 214 (2020).

\bibitem{bradiaelhami} B. Radi and A. El Hami, \textit{Advanced Numerical Methods with Matlab 2.} (John Wiley \& Sons, Inc., Hoboken, NJ, 2018).

\bibitem{yyuany19}Y. Du, L. Yuan, and Y. Wang, Commun. Comput. Phys. \textbf{26}, 768 (2019).

\end{thebibliography}
\end{document}